\documentclass[prc,onecolumn,floatfix,showpacs,preprintnumbers,amsmath,amssymb]{revtex4}
\usepackage{graphicx}
\usepackage{epsfig}
\usepackage{CJK}
\usepackage{amsmath}
\usepackage{supertabular}
\usepackage{textcomp}
\usepackage{threeparttable}
\usepackage{color}    
\usepackage{slashed}  
\usepackage{ulem}     
\usepackage{cancel}   
\usepackage{underscore}

\usepackage{hyperref}
\hypersetup{bookmarksnumbered,%
	colorlinks,%
	linkcolor=blue,%
	citecolor=blue,%
	plainpages=false,%
	pdfstartview=FitH}

\begin{document}
\normalem

  \title{Basic quantities of the Equation of State in isospin asymmetric nuclear matter}
  \author{Jie Liu$^{1}$, Chao Gao$^{1}$, Niu Wan$^{2}$, and Chang Xu$^{1,}$$\footnote{E-mail address: cxu@nju.edu.cn}$}
  \address{$^{1}$School of Physics, Nanjing University, Nanjing 210093, China\\
           $^{2}$School of Physics and Optoelectronics, South China University of Technology, Guangzhou 510641, China}
   \begin{abstract}
      Based on the Hugenholtz-Van Hove theorem, six basic quantities of the EoS in isospin asymmetric nuclear matter are expressed in terms of the nucleon kinetic energy $t(k)$, the isospin symmetric and asymmetric parts of the single-nucleon potentials $U_0(\rho,k)$ and $U_{\text{\text{sym,i}}}(\rho,k)$.
     The six basic quantities include the quadratic symmetry energy $E_{\text{sym,2}}(\rho)$, the quartic symmetry energy $E_{\text{sym,4}}(\rho)$, their corresponding density slopes $L_2(\rho)$ and $L_4(\rho)$, and the incompressibility coefficients $K_2(\rho)$ and $K_4(\rho)$.
     By using four types of well-known effective nucleon-nucleon interaction models, namely the BGBD, MDI, Skyrme, and Gogny forces, the density- and isospin-dependent properties of these basic quantities are systematically calculated and their values at the saturation density $\rho_0$ are explicitly given.
     The contributions to these quantities from $t(k)$, $U_0(\rho,k)$, and $U_{\text{sym,i}}(\rho,k)$ are also analyzed at the normal nuclear density $\rho_0$.
     It is clearly shown that the first-order asymmetric term $U_{\text{sym,1}}(\rho,k)$ (also known as the symmetry potential in Lane potential) plays a vital role in determining the density dependence of the quadratic symmetry energy $E_{\text{sym,2}}(\rho)$.
     It is also shown that the contributions from high-order asymmetric parts of the single-nucleon potentials ($U_{\text{sym,i}}(\rho,k)$ with $i>1$) cannot be neglected in the calculations of the other five basic quantities.
     Moreover, by analyzing the properties of asymmetric nuclear matter at the exact saturation density $\rho_{\text{sat}}(\delta)$, the corresponding quadratic incompressibility coefficient is found to have a simple empirical relation $K_{\text{sat,2}}=K_{2}(\rho_0)-4.14 L_2(\rho_0)$.
   \end{abstract}
    \maketitle

   \section{Introduction}

   Research on the isospin- and density-dependent properties of the equation of state (EoS) in isospin asymmetric nuclear matter is a longstanding issue in both nuclear physics and astrophysics \cite{PDan02,JML04,MBal16,CJJ21}.
   With respect to the exchange symmetry between protons and neutrons, the EoS for asymmetric nuclear matter can be expressed as an even series of isospin asymmetry $E(\rho,\delta) = E_0(\rho)+\overset{\sum}{_{i=2,4,\cdots}}{E_{\text{sym,i}}(\rho)\delta^{i}}$, in which the first term is the energy per nucleon in symmetric nuclear matter and the coefficients of the isospin-dependent terms are known as the $i$-th order symmetry energy $E_{\text{sym,i}}(\rho)=\frac{1}{i!}\frac{\partial^i E(\rho,\delta)}{\partial \delta^i}\mid_{\delta=0}$. In recent years, the EoS of nuclear matter has been extensively studied by
   (I) microscopic and phenomenological many-body approaches \cite{MB03,HY20,XLR21,JXU21};
   (II) the observables from heavy-ion reactions \cite{GC14,LWC14,JMDONG13,JXu09,GFWei20,OL09};
   (III) the astrophysical observations \cite{LiBA19,NBZhang-NST-18,YXU19}.
   For symmetric nuclear matter, the saturation density is constrained in a relatively narrow region $\rho_0=0.145\sim0.180$ fm${^{-3}}$ and the corresponding energy per nucleon $E_0(\rho_0)$ is approximately $-16$ MeV \cite{LiBA08}.
   The incompressibility coefficient $K_0(\rho_0)$ has a generally accepted value of $240\pm20$ MeV constrained by theoretical approaches and the giant monopole resonance data \cite{DHY99,JPB80,SS06}.
   In addition, the skewness $J_0(\rho_0)$ was recently found to have significant effects on the structures of neutron stars, but its value was scattered widely from $-800$ MeV to $400$ MeV \cite{BJCai17,NBZ18,WJX19}.
   For asymmetric nuclear matter, the value of the quadratic symmetry energy $E_{\text{sym,2}}(\rho_0)$ is constrained to be $31.7\pm3.2$ MeV \cite{LiBA13,MOer17}.
   However, its density slope and the incompressibility coefficient remain uncertain, \textit{i.e.} $L_2(\rho_0)=58.7\pm28.1$ MeV \cite{LiBA13,MOer17} and $K_{\text{sat,2}}=-550\pm100$ \cite{UGar07,TLi07,MLQ88}. It should be emphasized that, at both sub-saturation and supra-saturation densities, the quadratic symmetry energy is not well constrained, especially at supra-saturation densities \cite{ZGXIAO09,LWC07,CXu12CPL,NBZhang17}.
   The quartic symmetry energy $E_{\text{sym,4}}(\rho_0)$ is predicted to be less than $1$ MeV \cite{JP17,ZWL18,JMDONG18}.
   In contrast to the quadratic one, few studies have been conducted on the quartic density slope $L_4(\rho_0)$ and the corresponding incompressibility coefficient $K_4(\rho_0)$ \cite{CGB17}.

   In the present work, we perform a systematic analysis of six basic quantities in the EoS based on the Hugenholtz-Van Hove (HVH) theorem \cite{NMH58}, namely the $E_{\text{sym,2}}(\rho)$, $E_{\text{sym,4}}(\rho)$, $L_2(\rho)$, $L_4(\rho)$, $K_2(\rho)$, and $K_4(\rho)$.
   Among them, the properties of $E_{\text{sym,2}}(\rho)$, $E_{\text{sym,4}}(\rho)$, and their slopes $L_2(\rho)$ and $L_4(\rho)$ were re-analyzed \cite{CXu11,NW17,CXu14,NW16,MJ21}. The analytical expressions of the incompressibility coefficients $K_2(\rho)$ and $K_4(\rho)$ in terms of single-nucleon potentials are given for the first time.
   In the literature, there are various effective interaction models: transport models such as the Bombaci-Gale-Bertsch-Das Gupta (BGBD) interaction \cite{CG87,IB91,CBD03,JR04}, the isospin- and momentum-dependent MDI interaction \cite{CBD03,LiBA04a,LiBA04b,LWC05c}, and the Lanzhou quantum molecular dynamics (LQMD) model \cite{ZQFENG11,ZQFENG12,FZHang20}, the self-consistent mean-field approach including the zero-range momentum-dependent Skyrme interaction \cite{THRS59,DV72,YZWANG19}, the finite-range Gogny interaction \cite{DMB67,JD75,DG77} and the relativistic mean-field model \cite{BOGUTA77,FO13}, etc..
   The values of these quantities at the saturation density $\rho_0$ are calculated using two types of BGBD interactions, the MDI interactions with $x=-1$, $0$ and $1$, $16$ sets of the Skyrme interactions \cite{MD12,GSkI,KDE0v1,LNS,MSL0,AWS05,SKRA,SkT1,Skxs20,SQMC650,SV-sym32}, and $4$ sets of the Gogny interactions \cite{JFB91,FC08,SG09}.
   By taking the NRAPR Skyrme interaction as an example, we show the isospin- and density-dependent properties of the EoS for asymmetric nuclear matter explicitly. Meanwhile, for symmetric nuclear matter, $E_{0}(\rho)$, $K_0(\rho)$, and $J_0(\rho)$ are also analyzed in detail.
   It should be emphasized that the skewness $J_0(\rho_0)$ was recently found to be closely related to not only the maximum mass of neutron stars but also the radius of canonical neutron stars, and the calculations on $J_0(\rho)$ in the present work might be helpful in further determining the properties of neutron stars.
   In particular, the contributions from the high-order terms of the single-nucleon potential $U_{\text{sym,3}}(\rho,k)$ and $U_{\text{sym,4}}(\rho,k)$ to these basic quantities are evaluated in detail.

   The paper is organized as follows.
   In Sec. \uppercase\expandafter{\romannumeral2}, based on the HVH theorem, we express the basic quantities of the EoS in terms of the nucleon kinetic energy and the symmetric and asymmetric parts of the single-nucleon potential.
   The isospin-dependent saturation properties of the asymmetric nuclear matter are also discussed.
   In Sec. \uppercase\expandafter{\romannumeral3}, the calculated results by using four different effective interaction models are given. Finally, a summary is presented in Sec. \uppercase\expandafter{\romannumeral4}.

\section{Decomposition of basic quantities of EoS in terms of global optical potential components}

  \subsection{Basic quantities in the Equation of State of asymmetric nuclear matter}
     \begin{figure}[!h]
        \includegraphics[width=11cm]{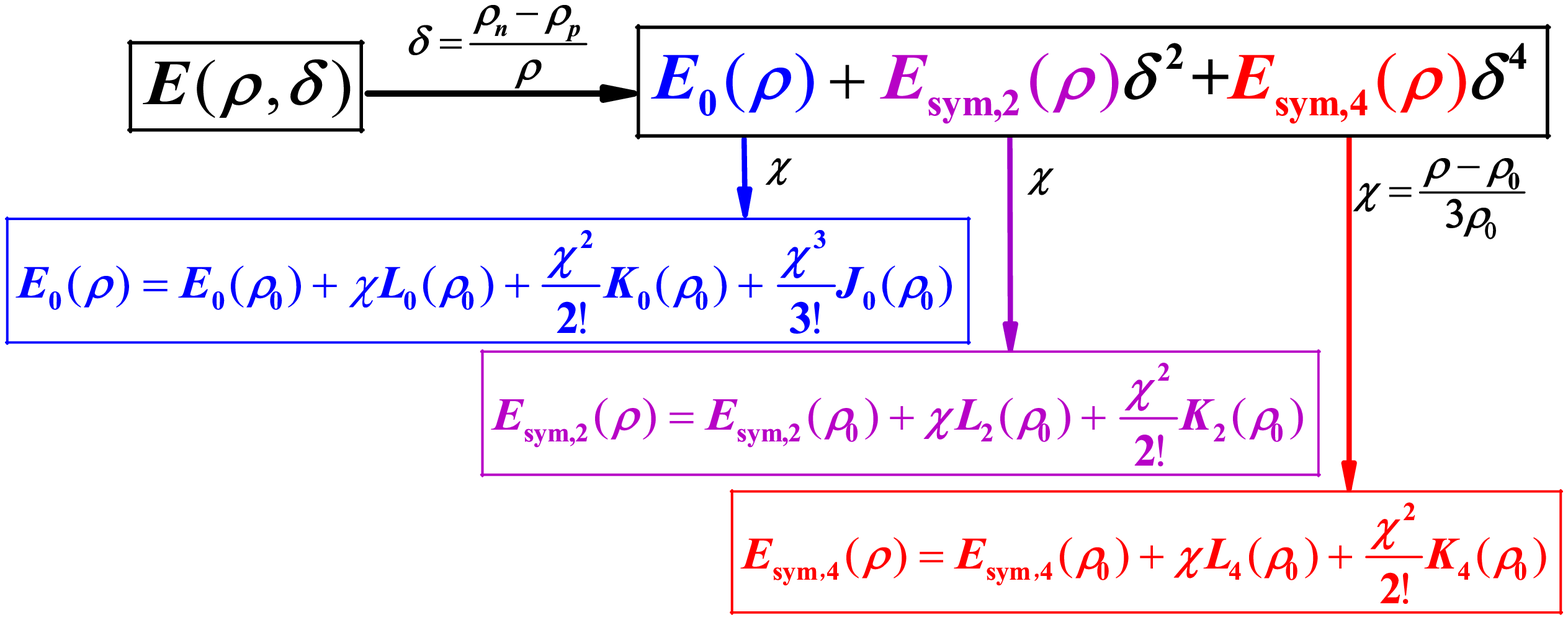}
        \caption{(Color online) The schematic diagram of basic quantities of the EoS in both isospin symmetric and asymmetric nuclear matter, including $E_0(\rho)$, $E_{\text{sym,2}}(\rho)$, $E_{\text{sym,4}}(\rho)$, $K_0(\rho_0)$, $J_0(\rho_0)$, $L_2(\rho_0)$, $K_2(\rho_0)$, $L_4(\rho_0)$, and $K_4(\rho_0)$.}
   \label{energy}
   \end{figure}
   For isospin asymmetric nuclear matter, the EoS can be expanded as a series of isospin asymmetry $\delta=(\rho_\text{n}-\rho_\text{p})/\rho$.
   If the high-order terms are neglected, it can be expressed as $E(\rho,\delta) = E_0(\rho)+E_{\text{sym,2}}(\rho)\delta^{2}+E_{\text{sym,4}}(\rho)\delta^{4}$ (see Fig. \ref{energy}).
   Each term can be further expanded around the saturation density of symmetric nuclear matter $\rho_0$ as a series of a dimensionless variable $\chi=\frac{\rho-\rho_0}{3\rho_0}$, which characterizes the deviations of the nuclear density $\rho$ from $\rho_0$.
   The density slope and incompressibility coefficient of the $i$-th order symmetry energy are defined as $L_i(\rho)=3\rho \frac{\partial E_{\text{sym,i}}(\rho) }{\partial \rho}$ and $K_i(\rho)=9\rho^2 \frac{\partial^2 E_{\text{sym,i}}(\rho) }{\partial \rho^2}$, respectively.
   The skewness of the EoS for symmetric nuclear matter is given by $J_0(\rho)=27\rho^3 \frac{\partial^3 E_{0}(\rho) }{\partial \rho^3}$.

   \subsection{The Hugenholtz-Van Hove (HVH) theorem and decomposition of basic quantities of asymmetric nuclear matter}

   Relating the Fermi energy $E_F$ and the energy per nucleon $E$, the general Hugenholtz-Van Hove (HVH) theorem can be written as \cite{NMH58}
   \begin{align}\label{HVH}
    	E_{F}=\frac{d \xi}{d \rho}=E+ \rho\frac{d E}{d \rho} = E+ \frac{P}{\rho},
   \end{align}
   where $\xi=\rho E$ and $P=\rho^2 \frac{\partial E}{\partial \rho}$ are the energy density and pressure of the fermion system at an absolute temperature of zero.
   Accordingly, the Fermi energies of neutrons and protons in asymmetric nuclear matter can be expressed as \cite{CXu11}
   \begin{subequations}\label{EF}
   	\begin{align}
   		t(k_{\text{F}}^{\text{n}})+U_\text{n}(\rho,\delta,k_{\text{F}}^{\text{n}})
   		=\frac{\partial \xi }{\partial \rho_\text{n}},  \label{EFn} \\
   		t(k_{\text{F}}^{\text{p}})+U_\text{p}(\rho,\delta,k_{\text{F}}^{\text{p}})
   		=\frac{\partial \xi }{\partial \rho_\text{p}},   \label{EFp}
   	\end{align}
   \end{subequations}
   where $t(k_{\text{F}}^{\text{n/p}})$ and $U_{\text{n/p}}(\rho,\delta,k_{\text{F}}^{\text{\text{n/p}}})$ are the kinetic energy and the single-nucleon potential of the neutron/proton with the Fermi momentum $k_{\text{F}}^{\text{\text{n/p}}}=k_{\text{F}}(1+\tau\delta)^{1/3}$.
   Furthermore, $U_{\text{\text{n/p}}}(\rho,\delta,k)$ can be expanded by a series of the isospin asymmetry $\delta$ as
\begin{align}
U_{\text{\text{n/p}}}(\rho,\delta,k)=
&U_0(\rho,k)+ U_{\text{sym,1}}(\rho,k) \tau\delta+U_{\text{sym,2}}(\rho,k) (\tau\delta)^2\nonumber\\
&+U_{\text{sym,3}}(\rho,k)(\tau\delta)^3 +U_{\text{sym,4}}(\rho,k)(\tau\delta)^4,
\end{align}
where $\tau=1$ is for the neutron and $\tau=-1$ for the proton, and $U_0(\rho,k)$ and $U_{\text{sym,i}}(\rho,k)$ are the symmetric and asymmetric parts, respectively.
   In particular, $U_0(\rho,k)$ and $U_{\text{sym,1}}(\rho,k)$ are called isoscalar and isovector (symmetry) potentials in the popular Lane potential \cite{AM62}.

   By subtracting Eq. (\ref{EFp}) from Eq. (\ref{EFn}), one can obtain
   \begin{align}
        [t(k_{\text{F}}^{\text{n}})-t(k_{\text{F}}^{\text{p}})]+[U_\text{n}(\rho,\delta,k_{\text{F}}^{\text{n}})-U_\text{p}(\rho,\delta,k_{\text{F}}^{\text{p}})]
        =\frac{\partial\xi }{\partial \rho_\text{n}} -\frac{\partial \xi }{\partial \rho_\text{p}}. \label{RLeq}
   \end{align}
   Expressing both sides of Eq. ({\ref{RLeq}}) in terms of $\delta$ and comparing the coefficients of $\delta$ and $\delta^3$, we can obtain the general expressions of the quadratic and quartic symmetry energies as
   \begin{subequations}\label{energies}
   	    \begin{align}
            E_{\text{sym,2}}(\rho) =& \frac{1}{6} \frac{\partial [t(k)+U_0(\rho,k)]}{\partial k}|_{k_{\text{F}}}k_{\text{F}}
   	                            + \frac{1}{2} U_{\text{sym,1}}(\rho,k_{\text{F}}), \label{E2}\\
            E_{\text{sym,4}}(\rho)
           	=& \frac{5}{324}\frac{\partial [t(k)+U_0(\rho,k)]}{\partial k}|_{k_{\text{F}}} k_{\text{F}}
           	- \frac{1}{108}\frac{\partial^2 [t(k)+U_0(\rho,k)]}{\partial k^2}|_{k_{\text{F}}} k_{\text{F}}^2    \nonumber   \\
           	& + \frac{1}{648} \frac{\partial^3 [t (k)+U_0(\rho,k)]}{\partial k^3}|_{k_{\text{F}}} k_{\text{F}}^3 \nonumber\\
   	        & - \frac{1}{36} \frac{ \partial U_{\text{sym,1}}(\rho,k)}{\partial k}|_{k_{\text{F}}} k_{\text{F}}
           	+ \frac{1}{72} \frac{ \partial^2 U_{\text{sym,1}}(\rho,k)}{\partial k^2}|_{k_{\text{F}}} k_{\text{F}}^2   \nonumber \\
           	& + \frac{1}{12} \frac{\partial U_{\text{sym,2}}(\rho,k)}{\partial k}|_{k_{\text{F}}} k_{\text{F}}
           	+ \frac{1}{4} U_{\text{sym,3}}(\rho,k_{\text{F}}). \label{E4}
   \end{align}
   \end{subequations}

   By adding Eqs. (\ref{EFn}) to (\ref{EFp}), expanding both sides of this summation in terms of $\delta$, and comparing the coefficients of $\delta^0$, we can obtain an important relationship between $E_0(\rho)$ and its density slope $L_0(\rho)$
   \begin{align}\label{EL}
      E_0(\rho)+ \rho \frac{\partial E_0(\rho) }{\partial \rho}=t(k_{\text{F}})+U_0(\rho,k_{\text{F}}),
   \end{align}
   where $L_0(\rho)$ is defined as $3\rho \frac{\partial E_0(\rho) }{\partial \rho}$ and can be rewritten as
   \begin{align}\label{ds0}
      L_0(\rho)=3[t(k_{\text{F}})+U_0(\rho,k_{\text{F}})]-3E_0(\rho).
   \end{align}
   Obviously, $E_0(\rho_0)=t(k_{\text{F}})+U_0(\rho_0,k_{\text{F}})$ and $E_0(\rho)$ can be calculated from the energy density of symmetric nuclear matter $\xi(\rho,\delta=0)$.
   Simultaneously, the general expressions of the density slopes $L_2(\rho)$ and $L_4(\rho)$ can also be given by comparing the coefficients of $\delta^2$ and $\delta^4$, namely,
    \begin{subequations}\label{ds}
   	    \begin{align}
   		 L_2(\rho)=&\frac{1}{6}\frac{\partial[t(k)+U_0(\rho,k)]}{\partial k}|_{k_{\text{F}}}k_{\text{F}} +\frac{1}{6}\frac{\partial^2
   			[t(k)+U_0(\rho,k)]}{\partial k^2}|_{k_{\text{F}}}k_{\text{F}}^2\nonumber \\
   		&+ \frac{\partial
   			U_{\text{sym,1}}(\rho,k)}{\partial k}|_{k_{\text{F}}}k_{\text{F}} + \frac{3}{2} U_{\text{sym,1}}(\rho,k_{\text{F}})+ 3U_{\text{sym,2}}(\rho,k_{\text{F}}),\label{ds2}\\
   		 L_4(\rho)=&\frac{5}{324}\frac{\partial[t(k)+U_0(\rho,k)]}{\partial k}|_{k_{\text{F}}}k_{\text{F}}-\frac{1}{324}\frac{\partial^2
   			[t(k)+U_0(\rho,k)]}{\partial k^2}|_{k_{\text{F}}}k_{\text{F}}^2\nonumber \\
   		&-\frac{1}{216}\frac{\partial^3
   			[t(k)+U_0(\rho,k)]}{\partial k^3}|_{k_{\text{F}}}k_{\text{F}}^3
   		+\frac{1}{648}\frac{\partial^4
   			[t(k)+U_0(\rho,k)]}{\partial k^4}|_{k_{\text{F}}}k_{\text{F}}^4\nonumber \\
   		&-\frac{7}{108}\frac{\partial
   			U_{\text{sym,1}}(\rho,k)}{\partial k}|_{k_{\text{F}}}k_{\text{F}} +\frac{1}{72}\frac{\partial^2
   			U_{\text{sym,1}}(\rho,k)}{\partial k^2 }|_{k_{\text{F}}}k_{\text{F}}^2+\frac{1}{54}\frac{\partial^3
   			U_{\text{sym,1}}(\rho,k)}{\partial k^3}|_{k_{\text{F}}}k_{\text{F}}^3 \nonumber \\
   		&+ \frac{5}{12}\frac{\partial
   			U_{\text{sym,2}}(\rho,k)}{\partial k}|_{k_{\text{F}}}k_{\text{F}}+\frac{1}{6}\frac{\partial^2
   			U_{\text{sym,2}}(\rho,k)}{\partial k^2}|_{k_{\text{F}}}k_{\text{F}}^2\nonumber \\
   		&+\frac{\partial
   			U_{\text{sym,3}}(\rho,k)}{\partial k}|_{k_{\text{F}}}k_{\text{F}}
   		 +\frac{9}{4}U_{\text{sym,3}}(\rho,k_{\text{F}})+3U_{\text{sym,4}}(\rho,k_{\text{F}}).\label{ds4}
   	\end{align}
   \end{subequations}	

    Taking the derivative of the summation of Eqs. (\ref{EFn}) and (\ref{EFp}) with respect to $\rho$ and comparing the coefficients, the incompressibility coefficients of $E_{0}(\rho)$, $E_{\text{sym,2}}(\rho)$, and $E_{\text{sym,4}}(\rho)$ are given as
   \begin{subequations}\label{K0K2K4}
   	\begin{align} K_0(\rho)=&9\rho\frac{\partial[t(k_{\text{F}})+U_0(\rho,k_{\text{F}})]}{\partial\rho}-18[t(k_{\text{F}})+U_0(\rho,k_{\text{F}})]+18E_0(\rho),\label{K0} \\
   K_2(\rho)=&-\frac{1}{3}\frac{\partial[t(k)+U_0(\rho,k)]}{\partial k}|_{k_{\text{F}}}k_{\text{F}}+\frac{1}{3}\frac{\partial^2
   		[t(k)+U_0(\rho,k)]}{\partial k^2}|_{k_{\text{F}}}k_{\text{F}}^2\nonumber \\
   	&-k_{\text{F}} \rho \frac{\partial}{\partial\rho}\frac{\partial[t(k)+U_0(\rho,k)]}{\partial k}|_{k_{\text{F}}}+\frac{1}{2}k_{\text{F}}^2 \rho\frac{\partial}{\partial\rho}\frac{\partial^2[t(k)+U_0(\rho,k)]}{\partial k^2}|_{k_{\text{F}}}\nonumber \\
   	&+\frac{\partial
   		U_{\text{sym,1}}(\rho,k)}{\partial k}|_{k_{\text{F}}}k_{\text{F}}+3k_{\text{F}}\rho \frac{\partial}{\partial\rho}\frac{\partial
   		U_{\text{sym,1}}(\rho,k)}{\partial k}|_{k_{\text{F}}}+9\rho \frac{\partial U_{\text{sym,2}}(\rho,k_{\text{F}})}{\partial\rho},\label{K2} \\
   K_4(\rho)=&-\frac{5}{162}\frac{\partial[t(k)+U_0(\rho,k)]}{\partial k}|_{k_{\text{F}}}k_{\text{F}}+\frac{23}{162}\frac{\partial^2
   		[t(k)+U_0(\rho,k)]}{\partial k^2}|_{k_{\text{F}}}k_{\text{F}}^2\nonumber \\
   	&-\frac{1}{12}\frac{\partial^3
   		[t(k)+U_0(\rho,k)]}{\partial k^3}|_{k_{\text{F}}}k_{\text{F}}^3
   	+\frac{5}{324}\frac{\partial^4
   		[t(k)+U_0(\rho,k)]}{\partial k^4}|_{k_{\text{F}}}k_{\text{F}}^4\nonumber \\
   	&-\frac{10}{27}k_{\text{F}} \rho \frac{\partial}{\partial\rho}\frac{\partial[t(k)+U_0(\rho,k)]}{\partial k}|_{k_{\text{F}}}+\frac{13}{54}k_{\text{F}}^2 \rho \frac{\partial}{\partial\rho}\frac{\partial^2[t(k)+U_0(\rho,k)]}{\partial k^2}|_{k_{\text{F}}}\nonumber \\
   	&-\frac{1}{18}k_{\text{F}}^3 \rho \frac{\partial}{\partial\rho}\frac{\partial^3[t(k)+U_0(\rho,k)]}{\partial k^3}|_{k_{\text{F}}}+\frac{1}{216}k_{\text{F}}^4 \rho \frac{\partial}{\partial\rho}\frac{\partial^4[t(k)+U_0(\rho,k)]}{\partial k^4}|_{k_{\text{F}}}
   	\nonumber \\
   	&-\frac{11}{54}\frac{\partial U_{\text{sym,1}}(\rho,k)}{\partial k}|_{k_{\text{F}}}k_{\text{F}} -\frac{5}{36}\frac{\partial^2
   		U_{\text{sym,1}}(\rho,k)}{\partial k^2}|_{k_{\text{F}}}k_{\text{F}}^2+\frac{1}{6}\frac{\partial^3
   		U_{\text{sym,1}}(\rho,k)}{\partial k^3}|_{k_{\text{F}}}k_{\text{F}}^3 \nonumber \\
   	&+\frac{5}{9}k_{\text{F}} \rho \frac{\partial}{\partial\rho}\frac{\partial U_{\text{sym,1}}(\rho,k)}{\partial k}|_{k_{\text{F}}}-\frac{1}{3}k_{\text{F}}^2 \rho \frac{\partial}{\partial\rho}\frac{\partial^2 U_{\text{sym,1}}(\rho,k)}{\partial k^2}|_{k_{\text{F}}}+\frac{1}{18}k_{\text{F}}^3 \rho \frac{\partial}{\partial\rho}\frac{\partial^3 U_{\text{sym,1}}(\rho,k)}{\partial k^3}|_{k_{\text{F}}}
   	\nonumber \\
   	&+\frac{13}{6}\frac{\partial
   		U_{\text{sym,2}}(\rho,k)}{\partial k}|_{k_{\text{F}}}k_{\text{F}}+\frac{4}{3}\frac{\partial^2
   		U_{\text{sym,2}}(\rho,k)}{\partial k^2}|_{k_{\text{F}}}k_{\text{F}}^2\nonumber \\
   	&-k_{\text{F}} \rho \frac{\partial}{\partial\rho}\frac{\partial U_{\text{sym,2}}(\rho,k)}{\partial k}|_{k_{\text{F}}}+\frac{1}{2}k_{\text{F}}^2 \rho \frac{\partial}{\partial\rho}\frac{\partial^2 U_{\text{sym,2}}(\rho,k)}{\partial k^2}|_{k_{\text{F}}}
   	\nonumber \\
   	&+\frac{27}{2}U_{\text{sym,3}}(\rho,k_{\text{F}})+7\frac{\partial
   		U_{\text{sym,3}}(\rho,k)}{\partial k}|_{k_{\text{F}}}k_{\text{F}}
   	+3 k_{\text{F}} \rho \frac{\partial}{\partial\rho}\frac{\partial U_{\text{sym,3}}(\rho,k)}{\partial k}|_{k_{\text{F}}}
   	\nonumber \\
   	&+18U_{\text{sym,4}}(\rho,k_{\text{F}})+9 \rho \frac{\partial U_{\text{sym,4}}(\rho,k_{\text{F}})}{\partial\rho}. \label{K4}
   	\end{align}
   \end{subequations}
   Similarly, taking the second derivative of Eq. (\ref{EL}) gives the skewness of $E_0(\rho)$ as follows
   	\begin{align}\label{J0}
      J_0(\rho)=  27\rho^2\frac{\partial^2[t(k_{\text{F}})+U_0(\rho,k_{\text{F}})]}{\partial\rho^2}
    	           -81\rho\frac{\partial[t(k_{\text{F}})+U_0(\rho,k_{\text{F}})]}{\partial\rho}
          +162[t(k_{\text{F}})+U_0(\rho,k_{\text{F}})]-162E_0(\rho).
	\end{align}

   \subsection{The exact saturation density $\rho_{\text{sat}}$ as a function of isospin asymmetry}

    For isospin asymmetric nuclear matter, the saturation density is different from that of the symmetric nuclear matter $\rho_0$. The former is defined as the exact saturation density and can be also written as a function of the isospin asymmetry $\delta$ \cite{LWC09}
    \begin{align}\label{rsat}
      \rho_{\text{sat}}(\delta)=\rho_{0}+\rho_{\text{sat,2}}\delta^2+\rho_{\text{sat,4}}\delta^4+O(\delta^6).
	\end{align}
    For symmetric nuclear matter with $\delta=0$, the $\rho_{\text{sat}}(\delta)$ is reduced to $\rho_0$. According to the property of the saturation point $\frac{\partial E(\rho,\delta)}{\partial \rho}|_{\rho_{\text{sat}} (\delta)}=0$ and expanding the EoS in terms of $\chi$, the exact saturation density can be expressed as
    \begin{align}\label{RS}
         \rho_{\text{sat}} (\delta)
         = \rho_0-\frac{3L_2(\rho_0)}{K_0(\rho_0)} \rho_0\cdot \delta^2+[\frac{3K_{2}(\rho_0)L_2(\rho_0)}{K_0(\rho_0)^2}
         -\frac{3L_4(\rho_0)}{K_0(\rho_0)}
          -\frac{3J_0(\rho_0)L_2^2(\rho_0)}{2K_0(\rho_0)^3}]  \rho_0\cdot \delta^4.
    \end{align}
    At the exact saturation density $\rho_{\text{sat}}(\delta)$, the energy per nucleon of asymmetric nuclear matter is given by
    \begin{align}\label{ES}
         E_{\text{sat}}(\delta)=E(\rho_{\text{sat}}(\delta),\delta)
         = & E_0(\rho_0)+E_{\text{sym,2}}(\rho_0)\delta^2+[E_{\text{sym,4}}(\rho_0)-\frac{L_2^2(\rho_0)}{2K_0(\rho_0)}]\delta^4
         \nonumber\\
         = & E_{sat,0}+E_{\text{sat,2}}\delta^2+E_{\text{sat,4}}\delta^4.
    \end{align}
    The corresponding incompressibility coefficient of the EoS is
    \begin{align}\label{KS}
         K_{\text{sat}}(\delta)= & 9\rho^2_{\text{sat}}(\delta)\frac{\partial^2 E(\rho,\delta)}{\partial^2 \rho}|_{\rho_{sat(\delta)}}  \nonumber\\
         = & K_0(\rho_0)+[K_{2}(\rho_0)-6L_2(\rho_0)-\frac{J_0(\rho_0)}{K_0(\rho_0)}L_2(\rho_0)]\delta^2+O[\delta^4]
          \nonumber\\
         = & K_{sat,0}+K_{\text{sat,2}}\delta^2+O[\delta^4].
    \end{align}
    It is clearly shown that the quartic symmetry energy at the exact saturation density is $E_{\text{sat,4}}=E_{\text{sym,4}}(\rho_0)-\frac{L_2^2(\rho_0)}{2K_0(\rho_0)}$, and the quadratic incompressibility coefficient is
    \begin{align}\label{KAS2}
         K_{\text{sat,2}}=K_{2}(\rho_0)-6L_2(\rho_0)-\frac{J_0(\rho_0)}{K_0(\rho_0)}L_2(\rho_0).
    \end{align}
    In previous studies \cite{LiBA08,MLQ88}, the $K_{\text{sat,2}}$ is approximated as $K_{\text{sat,2}}\rightarrow K_{\text{asy,2}} =K_{2}(\rho_0)-6L_2(\rho_0)$ by neglecting the $-\frac{J_0(\rho_0)}{K_0(\rho_0)}L_2(\rho_0)$ term for simplicity.
    We will discuss its effect on $K_{\text{sat,2}}$ in the following section.


   \section{Results and discussions}

\begin{figure}[!h]
     \includegraphics[width=16cm]{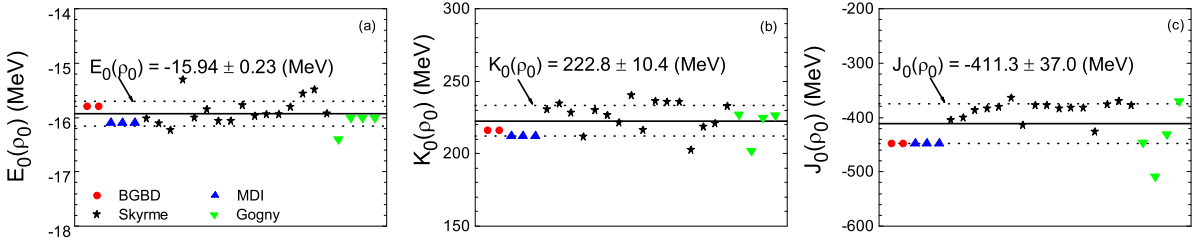}
     \caption{(Color online) Values of basic quantities $E_0(\rho_0)$, $K_0(\rho_0)$, and $J_0(\rho_0)$ for symmetric nuclear matter at $25$ parameter sets of the BGBD, MDI, Skyrme, and Gogny interactions. The solid and dashed lines represent the average values and their deviations, respectively.}
   \label{EKJ}
   \end{figure}

\begin{figure}[!h]
     \includegraphics[width=11cm]{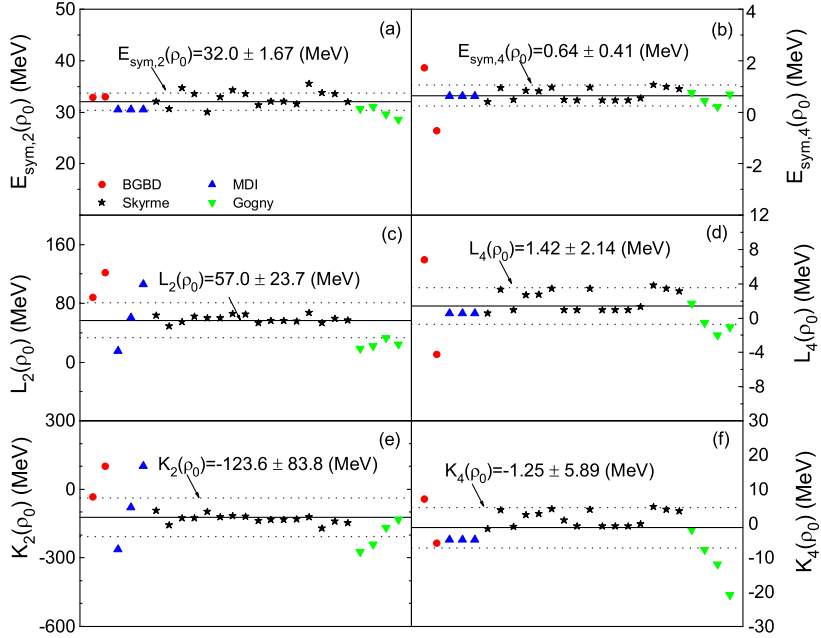}
     \caption{(Color online) Values of $E_{\text{sym,2}}(\rho_0)$, $L_2(\rho_0)$, $K_2(\rho_0)$,  $E_{\text{sym,4}}(\rho_0)$, $L_4(\rho_0)$, and $K_4(\rho_0)$ for asymmetric nuclear matter within $25$ parameter sets of four kinds of interactions. }
   \label{ELK6}
   \end{figure}

   We performed a systematic analysis of the basic quantities in the EoS of both symmetric and asymmetric nuclear matter at the saturation density $\rho_0$ by using $25$ interaction parameter sets, which include two BGBD interactions with different neutron-proton effective masses \cite{CG87,IB91,CBD03,JR04}, the MDI interaction with $x=-1$, $0$, and $1$ \cite{CBD03,LiBA04a,LiBA04b,LWC05c}, the $16$ Skyrme interactions \cite{MD12,GSkI,KDE0v1,LNS,MSL0,AWS05,SKRA,SkT1,Skxs20,SQMC650,SV-sym32}, and four Gogny interactions \cite{JFB91,FC08,SG09}.
   It is known that most of these interactions are fitted to the properties of finite nuclei, and the extrapolations to abnormal densities can be rather diverse.
   However, the comparison of a large number of results from different interactions could possibly provide useful information on the tendency of the density dependence of these basic quantities.
   Detailed numerical results from total $25$ interaction parameter sets are summarized in Table \ref{TableI}. The average values of the basic quantities in the EoS are also given.
   For comparison, we also list the constraints summarized in other studies (see the last row of Table \ref{TableI}).
   As shown in Table \ref{TableI}, the calculated values of $E_{0}(\rho_{0})$, $K_0(\rho_{0})$, $E_{\text{sym,2}}(\rho_{0})$, and $L_{2}(\rho_{0})$ are consistent with the constraints extracted from both theoretical calculations and experimental data \cite{LiBA08,SS06,LiBA13,MOer17}.
   Interestingly, the averaged $E_{\text{sym,4}}(\rho_{0})$ value is almost the same as the constraint in Ref. \cite{LWC09}.
   To further estimate the error bars of these basic quantities, all the calculated values in Table \ref{TableI} are plotted in Figs. \ref{EKJ} and \ref{ELK6}.
   It is seen from Fig. \ref{EKJ} that the data points of $E_{0}(\rho_{0})$ and $K_{0}(\rho_{0})$ are well constrained in a narrow range and the corresponding error bars are small.
   The error bar of skewness $J_0(\rho_0)=-411.3\pm37.0$ MeV is relatively large, especially for the Gogny interactions.
   It is also noted that the skewness, together with $K_{2}(\rho_0)$, has recently received much attention in the calculations of the neutron stars' maximum mass and the radius of canonical neutron stars \cite{LiBA19,NBZ18,WJX19}.
   The error bars of the high-order terms $L_{4}(\rho_0)$, $K_{2}(\rho_0)$ and $K_{4}(\rho_0)$ are also given, \textit{i.e.} $L_{4}(\rho_0)=1.42\pm2.14$ MeV, $K_{2}(\rho_0)=-123.6\pm83.8$ MeV, and $K_{4}(\rho_0)=-1.25\pm5.89$ MeV.
   In addition, for the MDI interaction, the $L_{2}(\rho_0)$ and $K_{2}(\rho_0)$ values with different spin(isospin)-dependent parameter $x$ are scattered over a wide range. This is because the different choices of parameter $x$ are to simulate very different density dependences of the symmetry energies at high densities \cite{CBD03,LiBA04a,LiBA04b}.

       \begin{table*}[!b]
    \caption{The saturation density $\rho _{0}$ (fm$^{-3}$) and basic quantities $E_{0}(\rho_{0})$, $K_0(\rho_{0})$, $J_{0}(\rho_{0})$, $E_{\text{sym,2}}(\rho_{0})$, $E_{\text{sym,4}}(\rho_{0})$, $L_{2}(\rho_{0})$, $L_4(\rho_{0})$, $K_{2}(\rho_{0})$, and $K_{4}(\rho_{0})$ for totally $25$ interaction sets in four kinds of interactions. The units of these quantities were MeV. In the last three rows, the averaged values and constraints in previous studies are shown. All interactions were taken from Ref. \cite{CG87,IB91,CBD03,JR04,LiBA04a,LiBA04b,LWC05c,MD12,GSkI,KDE0v1,LNS,MSL0,AWS05,SKRA,SkT1,Skxs20,SQMC650,SV-sym32,JFB91,FC08,SG09}.}
   \label{TableI}
   \begin{tabular*}{\textwidth}{@{\extracolsep{\fill}}lcccccccccccr}
   \hline\hline
   $$Force & $\rho_0$ & $E_0(\rho_0)$ &  $K_{0}(\rho_0)$ & $ J_{0}(\rho_0) $  & $E_{\text{sym,2}}(\rho_0)$ & $L_{2}(\rho_0)$  &  $K_{2}(\rho_0)$ & $E_{\text{sym,4}}(\rho_0)$  & $L_{4}(\rho_0)$ & $K_{4}(\rho_0)$ \\\hline
   $$ BGBD &&&&&&\\\hline
   $$ Case-1   & 0.160 &-15.8 & 215.9 & -447.5 & 32.9 &  87.9 & -32.7 & 1.72 & 6.82 & 7.14 \\
   $$ Case-2   & 0.160 &-15.8 & 215.9 & -447.5 & 33.0 & 121.8 & 101.0 & -0.73 & -4.26 & 7.14 \\\hline
   $$ MDI &&&&&&&&&&\\\hline
   $$  $x=1$   & 0.160 & -16.1 & 212.4 & -447.3 & 30.5 & 14.7  & -264.0 & 0.62 & 0.53 & -4.83 \\
   $$  $x=0$   & 0.160 & -16.1 & 212.4 & -447.3 & 30.5 & 60.2  & -81.7  & 0.62 & 0.53 & -4.83 \\
   $$ $x=-1$   & 0.160 & -16.1 & 212.4 & -447.3 & 30.5 & 105.8 & 100.6  & 0.62 & 0.53 & -4.83 \\\hline
   $$ Skyrme  &&&&&&&&&&  \\  \hline
   $$ GSKI    & 0.159 & -16.0 & 230.3 & -405.7 & 32.0 & 63.5 & -95.3  & 0.38 & 0.56 & -1.61 \\
   $$ GSKII   & 0.159 & -16.1 & 234.1 & -400.2 & 30.5 & 48.6 & -158.3 & 0.92 & 3.26 & 3.80  \\
   $$ KDE0v1  & 0.165 & -16.2 & 228.4 & -386.3 & 34.6 & 54.7 & -127.4 & 0.46 & 0.92 & -0.94 \\
   $$ LNS     & 0.175 & -15.3 & 211.5 & -384.0 & 33.5 & 61.5 & -127.7 & 0.82 & 2.67 & 2.44  \\
   $$ MSL0    & 0.160 & -16.0 & 230.0 & -380.3 & 30.0 & 60.0 & -99.3  & 0.81 & 2.70 & 2.66  \\
   $$ NRAPR   & 0.161 & -15.9 & 226.6 & -364.1 & 32.8 & 59.7 & -123.7 & 0.96 & 3.41 & 4.09  \\
   $$ Ska25s20& 0.161 & -16.1 & 221.5 & -415.0 & 34.2 & 65.1 & -118.2 & 0.46 & 0.93 & 0.88  \\
   $$ Ska35s20& 0.158 & -16.1 & 240.3 & -378.6 & 33.5 & 64.4 & -120.9 & 0.45 & 0.90 & -0.90 \\
   $$ SKRA    & 0.159 & -15.8 & 216.1 & -377.2 & 31.3 & 53.0 & -138.8 & 0.95 & 3.39 & 4.07  \\
   $$ SkT1    & 0.161 & -16.0 & 236.1 & -383.5 & 32.0 & 56.2 & -134.8 & 0.46 & 0.91 & -0.91 \\
   $$ SkT2    & 0.161 & -15.9 & 235.7 & -382.6 & 32.0 & 56.2 & -134.7 & 0.46 & 0.91 & -0.91 \\
   $$ SkT3    & 0.161 & -15.9 & 235.7 & -382.7 & 31.5 & 55.3 & -132.1 & 0.46 & 0.91 & -0.91 \\
   $$ Skxs20  & 0.162 & -15.8 & 202.4 & -426.5 & 35.5 & 67.1 & -122.5 & 0.53 & 1.27 & -0.22 \\
   $$ SQMC650 & 0.172 & -15.6 & 218.2 & -376.9 & 33.7 & 52.9 & -173.2 & 1.05 & 3.82 & 4.77  \\
   $$ SQMC700 & 0.171 & -15.5 & 220.7 & -369.9 & 33.5 & 59.1 & -140.8 & 0.97 & 3.44 & 4.03  \\
   $$ SV-sym32& 0.159 & -15.9 & 232.8 & -378.3 & 31.9 & 57.0 & -148.2 & 0.89 & 3.11 & 3.50  \\\hline
   $$ Gogny &&&&&&&&&\\\hline
   $$ D1      & 0.166 & -16.4 & 227.2 & -446.9 & 30.7 & 18.6 & -273.6 & 0.76 & 1.75 & -1.78 \\
   $$ D1S     & 0.163 & -16.0 & 201.8 & -508.4 & 31.1 & 22.5 & -241.0 & 0.44 & -0.51 & -7.56 \\
   $$ D1N     & 0.161 & -16.0 & 224.5 & -430.9 & 29.6 & 33.6 & -168.2 & 0.21 & -1.95 & -11.80 \\
   $$ D1M     & 0.165 & -16.0 & 226.2 & -466.9 & 28.6 & 24.8 & -133.3 & 0.69 & -1.05 & -20.81 \\\hline
   $$ Average  & 0.162 & -15.94 & 222.8 & -411.3 & 32.0 & 57.0 & -123.6 & 0.64 & 1.42 & -1.25 \\\hline
   $$ Constraint& & -16 & 240  && 31.7 & 58.7 && 0.62&&\\
   $$ Ref.   & & \cite{LiBA08} & \cite{SS06} &&   \cite{LiBA13,MOer17} & \cite{LiBA13,MOer17} &&  \cite{LWC09}&&\\
   \hline\hline
   \end{tabular*}
   \end{table*}


   \begin{figure}[t]
     \includegraphics[width=13.5cm]{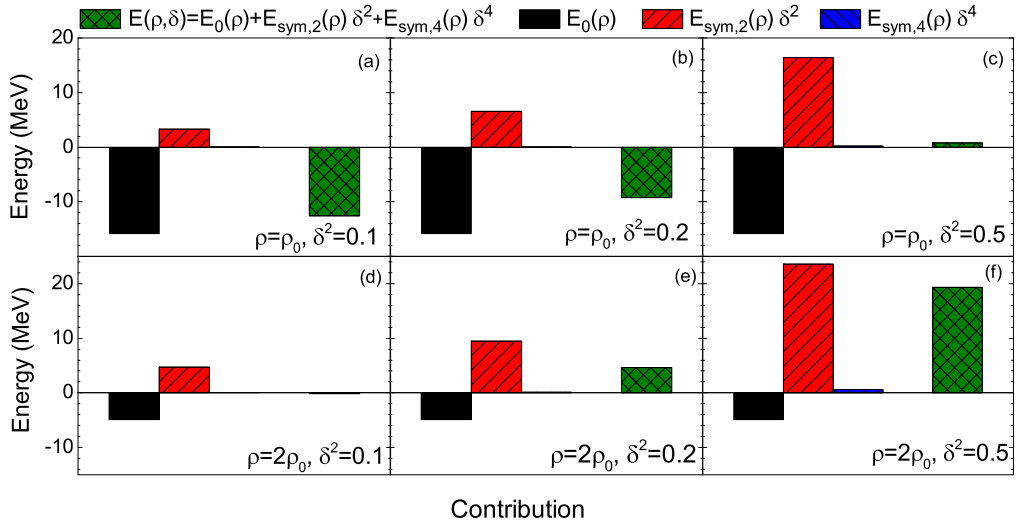}
     \caption{(Color online) The magnitudes of $E_{0}(\rho)$, $E_{\text{sym,2}}(\rho)\delta^2$, and $E_{\text{sym,4}}(\rho)\delta^4$ in the EoS at two different $\rho$ values and three different $\delta^2$ values.
     The NRAPR Skyrme interaction was applied.}
   \label{ERD}
   \end{figure}
   \begin{figure}[!h]
     \includegraphics[width=13.5cm]{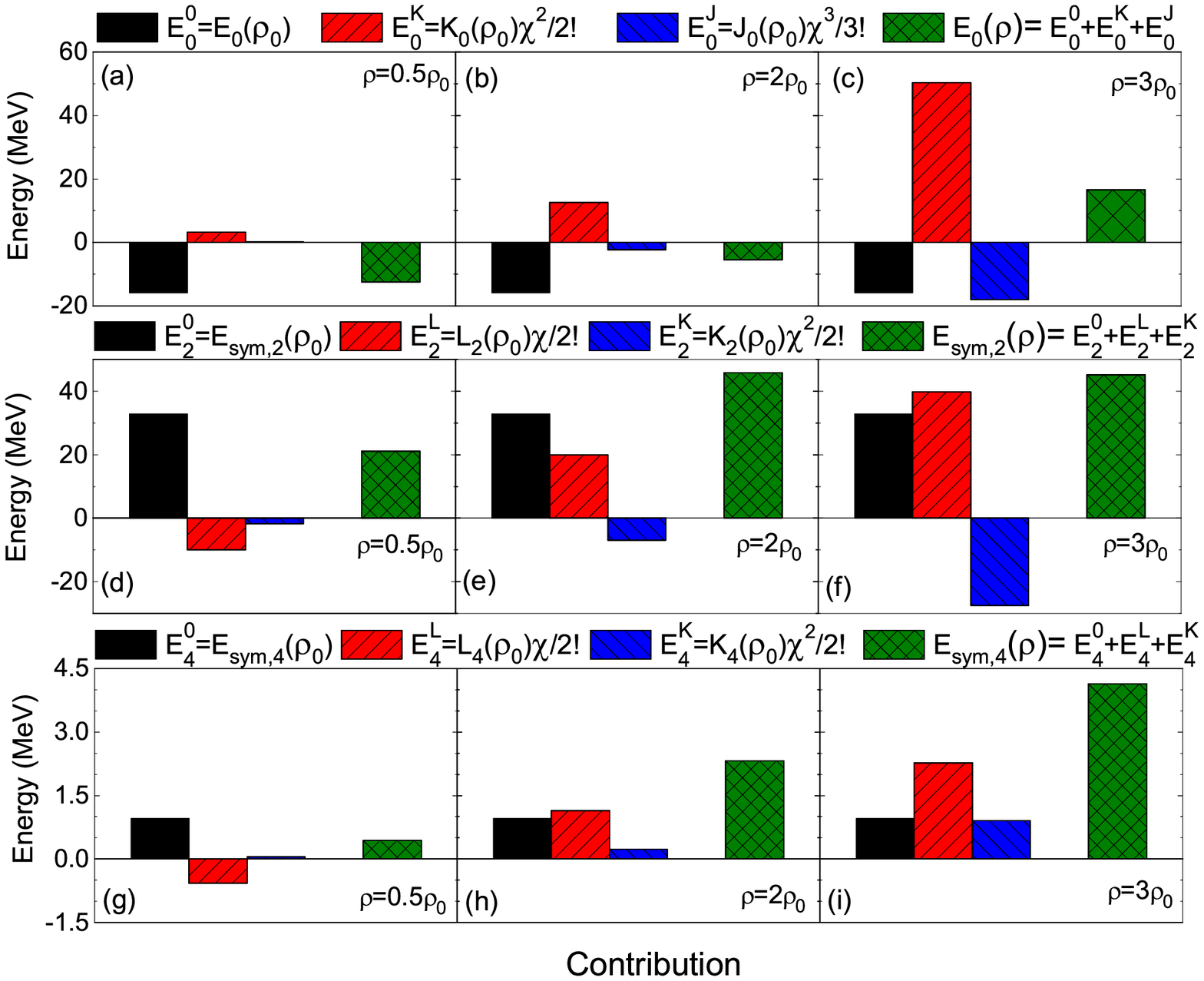}
     \caption{(Color online) The magnitude of each order in $E_{0}(\rho)$, $E_{\text{sym,2}}(\rho)$, and $E_{\text{sym,4}}(\rho)$ expressed by $E_{0}(\rho_0)$, $K_{0}(\rho_0)$ and $J_{0}(\rho_0)$, $E_{\text{sym,2}}(\rho_0)$, $L_{2}(\rho_0)$ and $K_{2}(\rho_0)$, and $E_{\text{sym,4}}(\rho_0)$, $L_{4}(\rho_0)$ and $K_{4}(\rho_0)$, respectively. The NRAPR Skyrme interaction was applied.}
   \label{Ex9}
   \end{figure}

    In Fig. \ref{ERD}, we show the magnitudes of the separated terms $E_{0}(\rho)$, $E_{\text{sym,2}}(\rho)\delta^2$, $E_{\text{sym,4}}(\rho)\delta^4$ as well as the total one $E(\rho,\delta)$ at two different densities ($\rho_0$ and $2\rho_0$) and three different isospin asymmetries ($\delta^2$=$0.1$, $0.2$ and $0.5$) by taking the NRAPR Skyrme interaction as an example.
    At the saturation density $\rho_0$ (see graphs (a)-(c)), the contribution of $E_{0}(\rho)$ to $E(\rho,\delta)$ is dominant.
    The contribution of $E_{\text{sym,2}}(\rho)\delta^2$ increases with an increase in isospin asymmetry $\delta$.
    It is also shown that the contribution from $E_{\text{sym,4}}(\rho)\delta^4$ is small and comes into play at large isospin asymmetry with $\delta^2=0.5$.
    At $2\rho_0$ (see graphs (d)-(f)), $E_{0}(\rho)$ contribution is suppressed compared with that at $\rho_0$, while $E_{\text{sym,2}}(\rho)\delta^2$ plays a more important role in the EoS, especially at $\delta^2=0.5$.
    It should also be noted that the $E_{\text{sym,4}}(\rho)$ contributes only at very high density and large isospin asymmetry. The magnitude of $E_{\text{sym,4}}(\rho)$ can significantly affect the calculation of the proton fraction in neutron stars at $\beta$-equilibrium \cite{JXu09,CXu11}.

   We further expand $E_{0}(\rho)$, $E_{\text{sym,2}}(\rho)$, and $E_{\text{sym,4}}(\rho)$ as a series of $\chi$ with their corresponding slopes and the incompressibility coefficients. In Fig. \ref{Ex9}, we depict the contributions from each term at different densities $0.5\rho_0$, $2\rho_0$ and $3\rho_0$.
   As can be seen from Fig. \ref{Ex9}, the first-order terms $E^0_0$ ($E_0(\rho_0)$), $E^0_2$ ($E_{\text{sym,2}}(\rho_0)$), and $E^0_4$ ($E_{\text{sym,4}}(\rho_0)$) contribute largely at all densities.
   $E_{0}^{\text{K}}$ and $E_{0}^{\text{J}}$ terms become increasingly important with increasing density.
   For $E_{\text{sym,2}}(\rho)$ and $E_{\text{sym,4}}(\rho)$ at $3\rho_0$, the contributions from their slopes ($E_2^{\text{L}}$ and $E_4^{\text{L}}$) and the incompressibility coefficients ($E_2^{\text{K}}$ and $E_4^{\text{K}}$) are much larger than those at $0.5\rho_0$ and $2\rho_0$.
   In particular, the $E_0^{\text{J}}$, $E_2^{\text{K}}$, and $E_4^{\text{K}}$ terms at $3\rho_0$ can be as important as the first-order terms.
   Thus, high-order terms should be considered when analyzing the properties of nuclear matter systems at high densities, such as the neutron stars.

   \begin{figure}[hbt]
    \includegraphics[width=13cm]{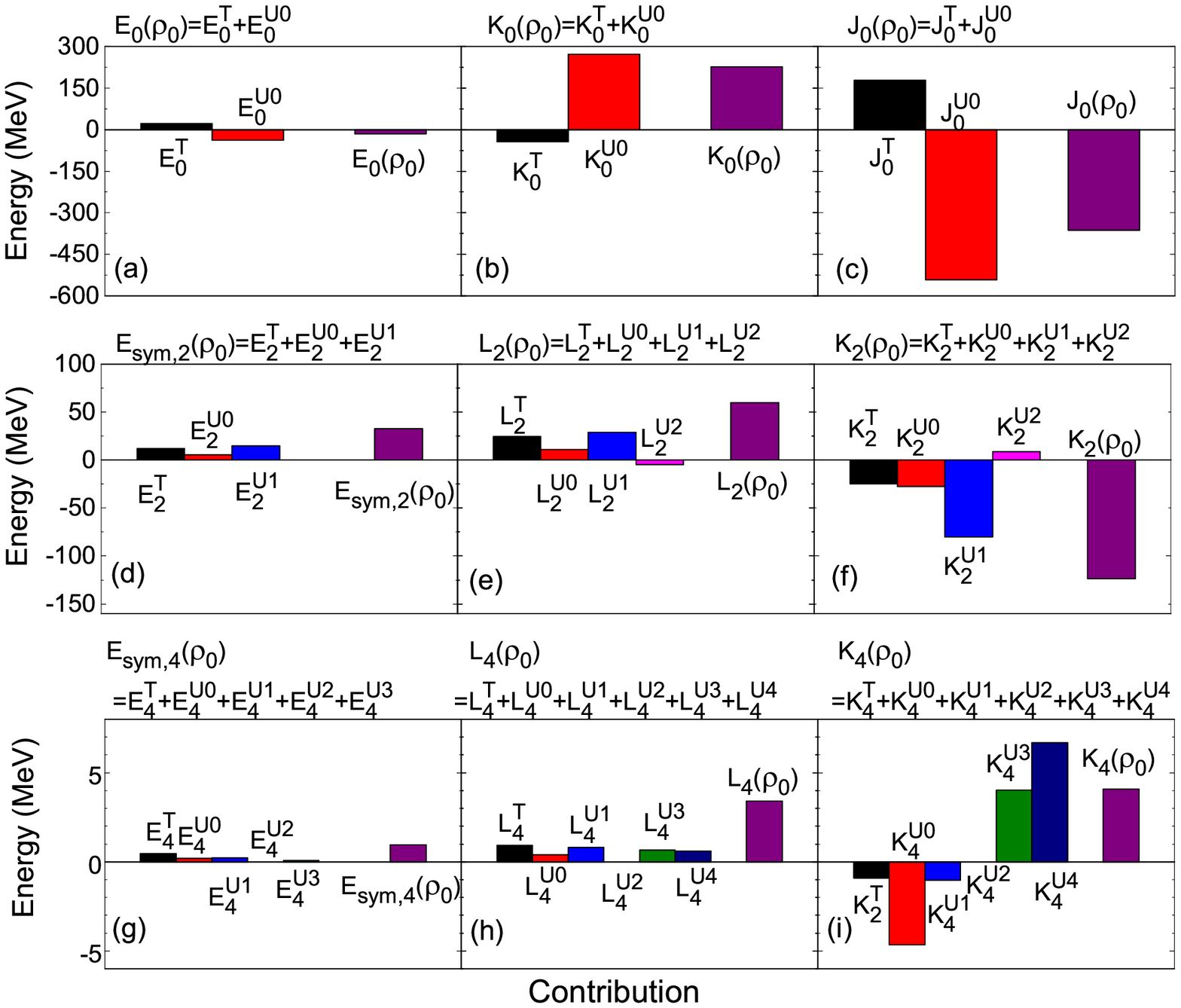}
    \caption{(Color online) The single-nucleon potential decomposition of $E_{0}(\rho_0)$, $K_{0}(\rho_0)$, $J_{0}(\rho_0)$, $E_{\text{sym,2}}(\rho_0)$, $L_{2}(\rho_0)$, $K_{2}(\rho_0)$, $E_{\text{sym,4}}(\rho_0)$, $L_{4}(\rho_0)$, and $K_{4}(\rho_0)$.
    The NRAPR Skyrme interaction was applied.}
   \label{value}
   \end{figure}
       \begin{figure}[hbt]
    \includegraphics[width=8cm]{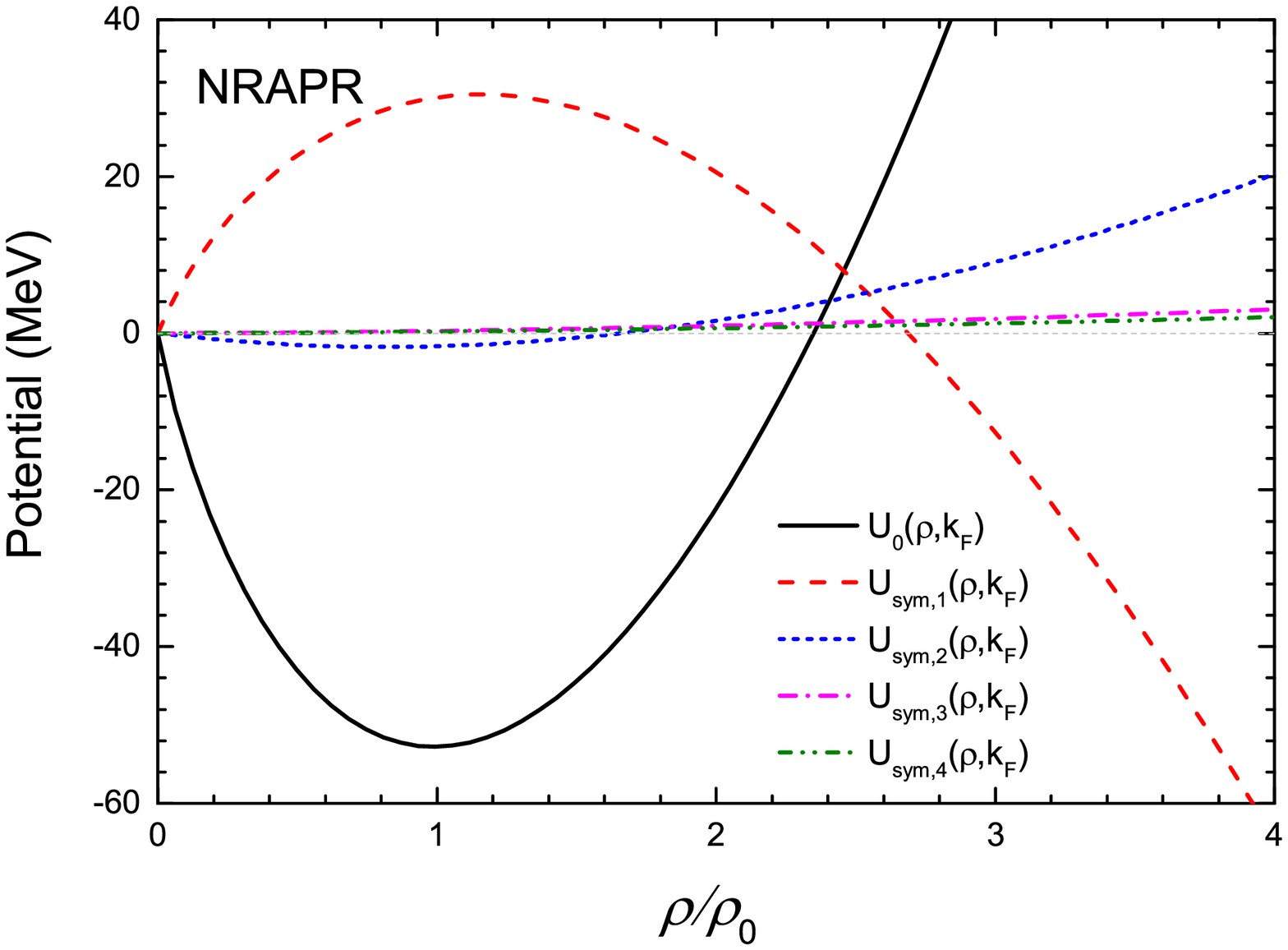}
    \caption{(Color online) The density-dependence of $U_{0}(\rho,k_{\text{F}})$, $U_{\text{sym,1}}(\rho,k_{\text{F}})$, $U_{\text{sym,2}}(\rho,k_{\text{F}})$, $U_{\text{sym,3}}(\rho,k_{\text{F}})$, and $U_{\text{sym,4}}(\rho,k_{\text{F}})$. The NRAPR Skyrme interaction was applied.}
    \label{potential}
    \end{figure}

    More interestingly, the basic quantities at the saturation density are decomposed into the kinetic energy $t(k)$ and the symmetric and asymmetric parts of the single-nucleon potential $U_0(\rho,k)$ and $U_{\text{sym,i}}(\rho,k)$.
    As shown in Fig. \ref{value}, the contributions from different terms $t(k)$, $U_0(\rho,k)$ and $U_{\text{sym,i}}(\rho,k) (i=1,2,3,4)$ are denoted by superscripts of $T$, $U0$, $U1$, $U2$, $U3$ and $U4$, respectively.
    It is clear that the $E_0(\rho_0)$, $K_0(\rho_0)$, and $J_0(\rho_0)$ are completely determined by $t(k)$ and $U_0(\rho,k)$. For other quantities, the contributions from the asymmetric parts $U_{\text{sym,1}}(\rho,k)$, $U_{\text{sym,2}}(\rho,k)$, $U_{\text{sym,3}}(\rho,k)$, and $U_{\text{sym,4}}(\rho,k)$ cannot be neglected.
    It is clearly shown that the first-order term $U_{\text{sym,1}}(\rho,k)$ contributes to all the six basic quantities.
    The second-order term $U_{\text{sym,2}}(\rho,k)$ does not contribute to $E_{\text{sym,2}}(\rho_0)$, but to its corresponding slope $L_{2}(\rho_0)$ and the incompressibility coefficient $K_{2}(\rho_0)$.
    In principle, the $U_{\text{sym,2}}(\rho,k)$ term should also contribute to the fourth-order terms $E_{\text{sym,4}}(\rho_0)$, $L_{4}(\rho_0)$, and $K_{4}(\rho_0)$, but for the Skyrme interaction, $U_{\text{sym,2}}(\rho,k)$ is not momentum-dependent and does not contribute.
    In addition, there are very few studies on the contributions of high-order terms $U_{\text{sym,3}}(\rho,k)$ and $U_{\text{sym,4}}(\rho,k)$ to the basic quantities.
    In Fig. \ref{potential}, we show the density-dependence of $U_{0}(\rho,k_{\text{F}})$, $U_{\text{sym,1}}(\rho,k_{\text{F}})$, $U_{\text{sym,2}}(\rho,k_{\text{F}})$, $U_{\text{sym,3}}(\rho,k_{\text{F}})$, and $U_{\text{sym,4}}(\rho,k_{\text{F}})$ at the Fermi momentum $k_{\text{F}}=(3\pi^2\rho/2)^{1/3}$ by using the NRAPR Skyrme interaction.
    It can be clearly seen from Fig. \ref{potential} that the magnitudes of $U_{0}(\rho,k_{\text{F}})$ and $U_{\text{sym,1}}(\rho,k_{\text{F}})$ are generally very large, while the ones of $U_{\text{sym,2}}(\rho,k_{\text{F}})$, $U_{\text{sym,3}}(\rho,k_{\text{F}})$ and $U_{\text{sym,4}}(\rho,k_{\text{F}})$ are very small but increase with the increasing density.
    Our results indicate that the $U_{\text{sym,3}}(\rho,k)$ and $U_{\text{sym,4}}(\rho,k)$ contributions should be taken into account for the fourth-order terms to understand the properties of asymmetric nuclear matter, especially for the cases with very large isospin asymmetries and high densities.


   \begin{table*}[!ht]
    \caption{The calculated values of expansion coefficients $\rho_0$ (fm$^{-3}$), $\rho_{\text{sat,2}}$ (fm$^{-3}$), $\rho_{\text{sat,4}}$ (fm$^{-3}$), the quartic symmetry energy $E_{\text{sat,4}}$ (MeV), the quadratic incompressibility coefficient $K_{\text{sat,2}}$ (MeV), and its two main components $K_{\text{asy,2}}$ (MeV) and $J_0(\rho_0)/K_0(\rho_0)$.
    In the last three rows, the averaged values and constraints in previous studies are shown. }
    \label{TableII}%
    \begin{tabular*}{\textwidth}{@{\extracolsep{\fill}}lcccccccr}
    \hline\hline
   Force &  $\rho_0$ & $\rho_{\text{sat,2}}$ & $\rho_{\text{sat,4}}$ & $E_{\text{sat,4}}$ & $K_{\text{asy,2}}$ &$K_{\text{sat,2}}$ & $J_0(\rho_0)/K_0(\rho_0)$  \\\hline
   $$ BGBD &&&&&&\\\hline
   $$ Case-1   & 0.160 & -0.195 & 0.038 & -16.17 & -560.1 & -377.9 & -2.07 \\
   $$ Case-2   & 0.160 & -0.271 & 0.295 & -35.11 & -630.0 & -377.5 & -2.07 \\\hline
   $$ MDI &&&&&&\\\hline
   $$   $x=1$  & 0.160 & -0.033 & -0.040 & 0.11 & -352.2  & -321.2 & -2.11 \\
   $$   $x=0$  & 0.160 & -0.136 & -0.013 & -7.91 & -442.9  & -316.1& -2.11 \\
   $$  $x=-1$  & 0.160 & -0.239 & 0.237 & -25.73 & -534.2 & -311.4 & -2.11 \\\hline
   $$ Skyrme &&&&&&\\\hline
   $$ GSKI    & 0.159 &  -0.131  &  -0.024  &  -8.36   &  -476.03  & -364.23  &  -1.76  \\
   $$ GSKII   & 0.159 &  -0.099  &  -0.056  &  -4.12   &  -450.04  & -366.94  &  -1.71  \\
   $$ KDE0v1  & 0.165 &  -0.119  &  -0.044  &  -6.09   &  -455.71  & -363.13  &  -1.69  \\
   $$ LNS     & 0.175 &  -0.153  &  -0.059  &  -8.12   &  -496.75  & -385.10  &  -1.82  \\
   $$ MSL0    & 0.160 &  -0.125  &  -0.033  &  -7.01   &  -459.33  & -360.11  &  -1.65  \\
   $$ NRAPR   & 0.161 &  -0.127  &  -0.050  &  -6.90   &  -481.82  & -385.91  &  -1.61  \\
   $$ Ska25s20& 0.161 &  -0.142  &  -0.039  &  -9.11   &  -508.89  & -386.89  &  -1.87  \\
   $$ Ska35s20& 0.158 &  -0.127  &  -0.039  &  -8.19   &  -507.47  & -405.95  &  -1.58  \\
   $$ SKRA    & 0.159 &  -0.117  &  -0.058  &  -5.55   &  -456.89  & -364.36  &  -1.75  \\
   $$ SkT1    & 0.161 &  -0.115  &  -0.045  &  -6.23   &  -471.90  & -380.66  &  -1.62  \\
   $$ SkT2    & 0.161 &  -0.115  &  -0.045  &  -6.23   &  -471.62  & -380.45  &  -1.62  \\
   $$ SkT3    & 0.161 &  -0.113  &  -0.044  &  -6.03   &  -463.93  & -374.14  &  -1.62  \\
   $$ Skxs20  & 0.162 &  -0.161  &  -0.044  &  -10.60  &  -525.16  & -383.74  &  -2.11  \\
   $$ SQMC650 & 0.172 &  -0.125  &  -0.082  &  -5.37   &  -490.78  & -399.34  &  -1.73  \\
   $$ SQMC700 & 0.171 &  -0.137  &  -0.065  &  -6.93   &  -495.14  & -396.16  &  -1.68  \\
   $$ SV-sym32& 0.159 &  -0.117  &  -0.057  &  -6.10   &  -490.44  & -397.74  &  -1.62  \\ \hline
   $$ Gogny &&&&&&\\\hline
   $$ D1      & 0.166 & -0.041 & -0.050 & 0.001 & -385.2 & -348.6 & -1.97 \\
   $$ D1S     & 0.163 & -0.055 & -0.056 & -0.81 & -376.0 & -319.3 & -2.52 \\
   $$ D1N     & 0.161 & -0.072 & -0.039 & -2.30 & -369.8 & -305.3 & -1.92 \\
   $$ D1M     & 0.165 & -0.054 & -0.023 & -0.67 & -282.1 & -230.9 & -2.06 \\\hline
   $$ Average  & 0.162 & -0.125 & -0.017 & -7.98 & -465.4 & -360.1 & -1.86 \\\hline
   $$ Constraint&  &  &  &  &     -500   & -370 / -550         &  \\
   $$   Ref.    &  &  &  &  & \cite{LWC07} & \cite{LWC09} / \cite{UGar07,TLi07}  &  \\
   \hline\hline
   \end{tabular*}
   \end{table*}

    \begin{figure}[!h]
     \includegraphics[width=14cm]{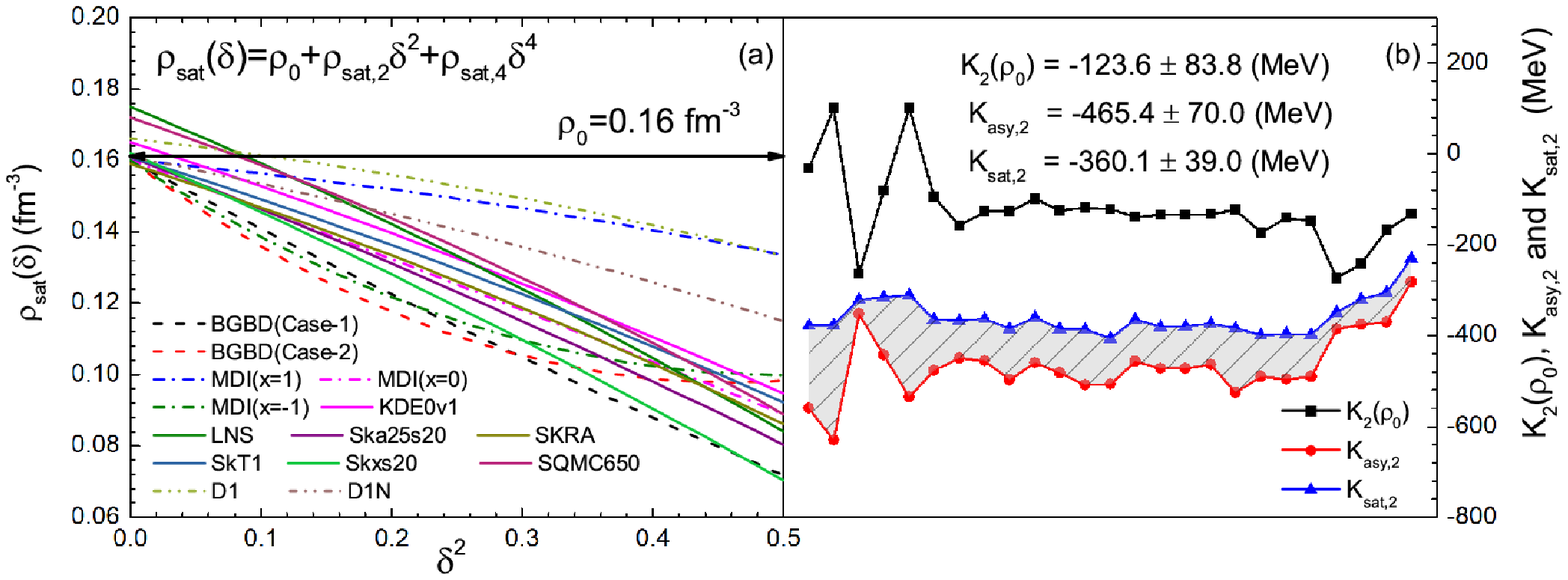}
     \caption{(Color online) The isospin-dependence of the exact saturation density $\rho_{\text{sat}}(\delta)$ within $14$ typical interaction parameter sets in graph (a) and the comparisons between the error bars of the quadratic incompressibility coefficients $K_2(\rho_0)$, $K_{\text{asy,2}}$, and $K_{\text{sat,2}}$ calculated by using $25$ interaction parameter sets in graph (b).}
   \label{re}
   \end{figure}

    By analyzing the isospin-dependence of the saturation properties of asymmetric nuclear matter, a number of important quantities are calculated using $25$ interaction parameter sets, and their numerical results as well as their averaged values are also listed in Table \ref{TableII}.
    For comparison, the constraints of $K_{\text{asy,2}}$ and $K_{\text{sat,2}}$ from other studies are listed in the last row of Table \ref{TableII}.
    It is shown that the second-order coefficient $\rho_{\text{sat,2}}$, one of the most important isospin-dependent parts of $\rho_{\text{sat}}(\delta)$, has a negative value in all cases, and the fourth-order coefficient $\rho_{\text{sat,4}}$ also has a negative value for the Skyrme and Gogny interactions. This means that in most cases, the saturation density of asymmetric nuclear matter is lower than that of symmetric nuclear matter, especially at larger isospin asymmetry $\delta$ (see graph (a) of Fig. \ref{re}).
    For the BGBD interaction (Case-2), the calculated value of $\rho_{\text{sat,4}}$ is positive and relatively large.
    According to the relationship in Eq. (\ref{rsat}), this would lead to a higher saturation density of asymmetric nuclear matter than that of symmetric nuclear matter with the isospin asymmetry $\delta$ close to unit.
    For asymmetric nuclear matter at $\rho_{\text{sat}}(\delta)$, the corresponding $E_{\text{sat,4}}$ values are rather diverse, which is considered to be important for the proton fraction in neutron stars.

    As shown in graph (b) of Fig. \ref{re}, the results of $K_2(\rho_0)$, $K_{\text{asy,2}}$, and $K_{\text{sat,2}}$ are given and their values are constrained to be $K_{2}=-123.6\pm83.8$ MeV, $K_{\text{asy,2}}=-465.4\pm70.0$ MeV, and $K_{\text{sat,2}}=-360.1\pm39.0$ MeV, respectively.
    The averaged $K_{\text{asy,2}}$ value is close to the previous theoretical constraint $-500\pm50$ MeV given in Ref. \cite{LWC07} if the error bar is considered.
    In Table II, there are two previous constraints for $K_{\text{sat,2}}$. One is $K_{\text{sat,2}}=-370\pm120$ MeV from a modified Skyrme-like (MSL) model \cite{LWC09}, and the other is $-550\pm100$ MeV by analyzing the measured data of the isotopic dependence of the giant monopole resonance (GMR) in the even-A Sn isotopes \cite{UGar07,TLi07}.
    Compared with these previous studies, it is clear that the $K_{\text{asy,2}}$ and $K_{\text{sat,2}}$ values remain uncertain and require more data to further constrain their values.
    In addition, as mentioned before, the term $-\frac{J_0(\rho_0)}{K_0(\rho_0)}L_2(\rho_0)$ in Eq. (\ref{KAS2}) is typically ignored for simplicity.
    However, it is clearly shown in Fig. \ref{re}(b) that the contribution of this term is non-negligible.
    In the present work, we include the contribution of this high-order term, and the ratio $J_0(\rho_0)/K_0(\rho_0)$ is constrained in the range of $-1.86\pm0.23$.
    Finally we obtain a simple relation for $K_{\text{sat,2}}$
       \begin{equation}
         K_{\text{\text{sat,2}}}=K_2(\rho_0)-4.14L_2(\rho_0) \label{KSATs}.	
	   \end{equation}
    With the averaged results $L_2(\rho_0)=57.0$ MeV and $K_2(\rho_0)=-123.6$ MeV, the calculated value $K_{\text{\text{sat,2}}}=-359.6$ MeV is in good agreement with the average value of $-360.1\pm39.0$ MeV from the $25$ interaction sets. This simple empirical relation could be useful for estimating the value of $K_{\text{\text{sat,2}}}$ for asymmetric nuclear matter.


   \section{SUMMARY}

     Based on the Hugenholtz-Van Hove theorem, the general expressions for the six basic quantities of EoS are expanded in terms of the kinetic energy $t(k)$, the symmetric and asymmetric parts of global optical potential $U_0(\rho,k)$ and $U_{\text{sym,i}}(\rho,k)$. The analytical expressions of coefficients $K_2(\rho)$ and $K_4(\rho)$ are given for the fist time.
     By using $25$ kinds of interaction sets, the values of these quantities were systematically calculated at the saturation density $\rho_0$.
     It is emphasized that there are very few studies on quantities $L_4(\rho_0)$, $K_2(\rho_0)$ and $K_4(\rho_0)$ and their average values from a total of $25$ interaction sets are $L_4(\rho_0)=1.42\pm2.14$ MeV, $K_2(\rho_0)=-123.6\pm83.8$ MeV, and $K_4(\rho_0)=-1.25\pm5.89$ MeV, respectively.
     The averaged values of the other quantities weere consistent with those of previous studies.
     Furthermore, the different contributions of the kinetic term, the isoscalar and isovector potentials to these basic quantities were systematically analyzed at the saturation density.
     It is clearly shown that $t(k_{\text{F}})$ and $U_0(\rho,k_{\text{F}})$ play vital roles in determining the EoS of both symmetric and asymmetric nuclear matter.
     For asymmetric nuclear matter, the $U_{\text{sym,1}}(\rho,k)$ contributes to all the quantities, whereas $U_{\text{sym,2}}(\rho,k)$ does not contribute to $E_{\text{sym,2}}(\rho_0)$, but contributes to the second-order terms $L_{2}(\rho_0)$ and $K_{2}(\rho_0)$ as well as the fourth-order terms $E_{\text{sym,4}}(\rho_0)$, $L_{4}(\rho_0)$, and $K_{4}(\rho_0)$.
     The contribution from $U_{\text{sym,3}}(\rho,k)$ cannot be neglected for $E_{\text{sym,4}}(\rho_0)$, $L_{4}(\rho_0)$, and $K_{4}(\rho_0)$.
     $U_{\text{sym,4}}(\rho,k)$ should also be included in the calculations for $L_{4}(\rho_0)$ and $K_{4}(\rho_0)$.
     In addition, the quadratic incompressibility coefficient at $\rho_{\text{sat}}(\delta)$ is found to have a simple empirical relation $K_{\text{sat,2}}=K_{2}(\rho_0)-4.14 L_2(\rho_0)$ based on the present analysis.

   \section*{ACKNOWLEDGEMENTS}

   This work is supported by the National Natural Science Foundation of China (Grant No. 11822503).

   \end{document}